\documentclass[apj]{emulateapj}
\usepackage{graphicx}
\usepackage{natbib}

\shorttitle{The Biggest Explosions in the Universe}
\shortauthors{Johnson, Whalen, Even, Fryer, Heger, Smidt and Chen}

\begin{document}

\title{The Biggest Explosions in the Universe}

\author{Jarrett L. Johnson\altaffilmark{1}, Daniel
  J. Whalen\altaffilmark{1,2}, Wesley
  Even\altaffilmark{3}, Chris L. Fryer\altaffilmark{3}, \\
 Alex Heger\altaffilmark{4}, Joseph Smidt\altaffilmark{1} 
  and Ke-Jung Chen\altaffilmark{5}}
\affil{$^{1}$Nuclear and Particle Physics, Astrophysics and Cosmology
  Group (T-2), \\
Thermonuclear Applications Physics Group (XTD-6), \\
Los Alamos National Laboratory, Los Alamos, NM 87545; jlj@lanl.gov}
\affil{$^{2}$Universit\"{a}t Heidelberg, Zentrum f\"{u}r Astronomie, 
Institut f\"{u}r Theoretische Astrophysik, \\
Albert-Ueberle-Str. 2, 69120 Heidelberg, Germany}
\affil{$^{3}$Computational Physics and Methods Group (CCS-2), \\ 
Los Alamos National Laboratory, Los Alamos, NM 87545}
\affil{$^{4}$Monash Centre for Astrophysics, Monash University,
  Victoria, 3800, Australia}
\affil{$^{5}$School of Physics and Astronomy, University of Minnesota,
Minneapolis, MN  55455}

%\topmargin+1.5cm

\begin{abstract}

Supermassive primordial stars are expected to form in a small fraction
of massive protogalaxies in the early universe, and are generally
conceived of as the progenitors of the seeds of supermassive black
holes (BHs).  Supermassive stars with masses of
$\sim55,000\,$M$_{\odot}$, however, have been found to
explode and completely disrupt in a supernova (SN) with an energy of
up to $\sim10^{55}\,$erg instead of collapsing to a BH.  Such events,
$\sim10,000$ times more energetic than typical SNe today, would be
among the biggest explosions in the history of the universe.  Here we
present a simulation of such a SN in two stages. Using the {\sc RAGE}
radiation hydrodynamics code we first evolve the explosion from an
early stage through the breakout of the shock from the surface of the
star until the blast wave has propagated out to several parsecs from
the explosion site, which lies deep within an atomic cooling dark
matter (DM) halo at $z\simeq15$.  Then, using the {\sc GADGET}
cosmological hydrodynamics code we evolve the explosion out to several
kiloparsecs from the explosion site, far into the low-density
intergalactic medium.  The host DM halo, with a total mass of $4\times
10^7\,$M$_{\odot}$, much more massive than typical primordial
star-forming halos, is completely evacuated of high density gas after
$\la 10\,$Myr, although dense metal-enriched gas recollapses into the
halo, where it will likely form second-generation stars with metallicities  of 
$\simeq 0.05\,$Z$_{\odot}$ after $\ga70\,$Myr.  The chemical signature
of supermassive star explosions may be found in such long-lived second-generation stars today.
 \end{abstract}


\section{Introduction}

Recently, there has been renewed interest in the long-standing
theoretical possibility that supermassive stars (SMSs), with masses of
$10^4$--$10^6\,$M$_{\odot}$ inhabited the early universe (see e.g.,
Volonteri 2012) and their possible fates (e.g., Iben 1963; Fowler \&
Hoyle 1964; Appenzeller \& Fricke 1972; Shapiro \& Teukolsky 1979;
Bond et al.\ 1984; Fuller et al.\ 1986).

One of the main motivations for their study comes from observations of
quasars at $z$ $\simeq 6$--$7$ which are inferred to be powered by
black holes (BHs) with masses exceeding $10^9\,$M$_{\odot}$ (e.g.,
Willott et al.\ 2003; Fan et al.\ 2006; Mortlock et al.\ 2011).  Given
the short time ($< 800\,$Myr) available for such massive BHs to grow
via accretion from their initial 'seed' masses, as derived from the
most recent cosmological parameters inferred by e.g., the {\it
  Wilkinson Microwave Anisotropy Probe} (Komatsu et
al.\ 2011),\footnote{Adopting the cosmological parameters reported
  recently by the {\it Planck} Collaboration (2013) yields a similar
  time available for seed growth.}  and the suppression of BH growth
due to the strong radiative feedback from both stars (e.g., Whalen et
al.\ 2004; Wise \& Abel 2007; O'Shea \& Norman 2008) and the BHs
themselves (e.g., Pelupessy et al.\ 2007; Alvarez et al.\ 2009;
Milosavljevi{\' c} et al.\ 2009; Jeon et al.\ 2012; Park \& Ricotti
2012), it now appears more likely than ever that the seeds of the most
massive early BHs must have been quite massive themselves (e.g., $\ga
10^5\,$M$_{\odot}$; see Johnson et al.\ 2012a; also e.g., Shapiro 2005;
Volonteri \& Rees 2006; Natarajan \& Volonteri 2012).  Whereas the
majority of the first, Population (Pop) III stars may have had masses
of $\sim 20$--$500\,$M$_{\odot}$ (e.g., Abel et al.\ 2002; Bromm \& Larson
2004; Yoshida et al.\ 2008; Greif et al.\ 2011), the best candidates for
the seeds of SMBHs are thus much more massive (and rare) supermassive
primordial stars.

An additional, and independent, reason to consider SMSs in the early
universe is that the conditions required for their formation are now
thought to be realized much more often than was previously assumed.
The most widely discussed avenue for the formation of SMSs is via the
direct gravitational collapse of hot ($\simeq10^4\,$K) primordial gas
in so-called atomic cooling dark matter (DM) halos at $z\ga 10$ (e.g.,
Bromm \& Loeb 2003; Begelman et al.\ 2006; Lodato \& Natarajan 2006;
Spaans \& Silk 2006; Regan \& Haehnelt 2009; Choi et al.\ 2013; Latif
et al.\ 2013a,b).\footnote{We note that other formation mechanisms
  stemming from shocks (Inayoshi et al.\ 2012) and magnetic fields
  (Sethi et al.\ 2010) have also been suggested.}  In this scenario,
the gas in the protogalaxy remains at the virial temperature of $\sim$
$10^4\,$K because H$_{\rm 2}$ molecules have been photodissociated by
the Lyman-Werner (LW) background, leading to the rapid formation of
SMSs via the accretion of gas at rates $\sim 10^2$--$10^3$ times
higher than in the formation of most Pop~III stars from H$_{\rm
  2}$-cooled gas.  The flux of radiation required to keep the gas
H$_{\rm 2}$-free depends on its spectrum, with lower fluxes required
if it is produced by metal-enriched stars instead of Pop~III stars
(e.g., Shang et al.\ 2010).  Recent work by independent groups has
shown that Pop~II star-forming galaxies in the early universe are able
to produce sufficient H$_{\rm 2}$-dissociating radiation to prevent
the cooling of primordial gas in a substantial fraction of atomic
cooling halos, thereby leading to the seeding of these halos with SMSs
that can collapse into BHs (see Dijkstra et al.\ 2008; Agarwal et
al.\ 2012; Petri et al.\ 2012; Johnson et al.\ 2013).  Indeed, Agarwal et
al.\ (2012, 2013) find that a large fraction of the SMBHs in the
centers of galaxies today may have been seeded by SMSs.  Strengthening
these conclusions are other recent results which suggest that lower LW
fluxes may be required for the formation of SMSs, due to a reduced
role of H$_{\rm 2}$ self-shielding (Wolcott-Green et al.\ 2011) and the
presence of significant turbulence or magnetic fields (Van Borm \&
Spaans 2013).

Complementary studies have been undertaken to understand the growth
and evolution of SMSs, as well. Modeling the growth of accreting
protostars with masses up to $\simeq 10^3\,$M$_{\odot}$, Hosokawa et
al.\ (2012) have shown that they emit little high energy radiation that
could halt their continued accretion, and Inayoshi et al.\ (2013) have
shown that pulsational instabilities are likewise unable to halt their
growth.  Johnson et al.\ (2012b) modeled the growth of SMSs to much
higher masses and showed that, even if they are able to emit the
copious ionizing radiation characteristic of main sequence Pop~III
stars, which may occur once the accretion rate becomes sufficiently
low (Schleicher et al.\ 2013), radiative feedback is not able to stop
their growth up to at least $\sim 10^5\,$M$_{\odot}$.  At the highest
accretion rates expected for these objects ($\ga 1\,$M$_{\odot}$
yr$^{-1}$; Wise et al.\ 2008; Shang et al.\ 2010; Johnson et al.\ 2011),
the masses of primordial SMSs are only limited by the $\la 4\, $Myr that
they have to accrete gas before they collapse to BHs (Begelman 2010).

Whereas the majority of SMSs are expected to collapse to black holes
with little or no associated explosion (e.g., Fryer \& Heger 2011; see
also Fuller \& Shi 1998; Linke et al.\ 2001), it is possible that some
fraction instead explode as extremely energetic supernovae (SNe; e.g.,
Fuller et al.\ 1986; Montero et al.\ 2012; Whalen et al.\ 2012a, 2013d).  In
particular, Heger et al.\ (2013) have found from stellar evolution
calculations including post-Newtonian corrections to gravity that SMSs
with masses in a narrow range around $\simeq 55,000\,$M$_{\odot}$ end
their lives as extraordinarily luminous SNe.\footnote{These SNe would
  appear much brighter than other types of Pop~III SNe (e.g.,
  Scannapieco et al.\ 2005; Hummel et al.\ 2012; Pan et al.\ 2012; Tanaka
  et al.\ 2012, 2013; Whalen et al.\ 2012b, 2013a,b,c; de Souza et
  al. 2013).}  With energies of
almost $10^{55}\,$erg, these thermonuclear explosions are among the
most energetic in the history of the universe.

Here we expand on the radiation hydrodynamics simulations presented by
Whalen et al.\ (2012a) and simulate the long-term evolution of a SMS SN
in its cosmological environment, in order to show how these gargantuan  
explosions impact both the formation of the first galaxies and the
chemical signature of the first stars.  In the next section, we
describe the multi-scale simulations that we have carried out to model 
the evolution of the explosion from its breakout from the surface of
the star to the propagation of the blast wave into the intergalactic
medium (IGM).  In Section~3, we present our results on the energetics
and dynamics of the explosion, as well as on metal enrichment and
second-generation star formation. In Section~4, we conclude with a
brief discussion of our results.

\begin{figure}
  \begin{center}
    \leavevmode
      \epsfxsize=8.5cm\epsfbox{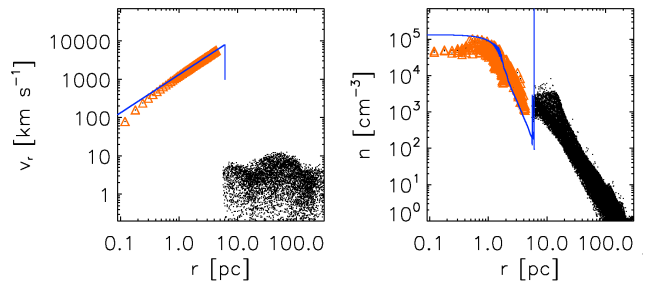}
       \caption{The initial conditions for the {\sc GADGET} 3-D
         cosmological simulation.  The blue curves show the radial
         velocity ({\it left}) and hydrogen number density ({\it
           right}) profiles resulting from the 1-D {\sc RAGE}
         simulation of the early evolution of the SMS SN remnant, up
         to the time when the shock extends out to 6 pc from the
         explosion site.  The orange triangles denote the {\sc GADGET}
         SPH particles representing the SN ejecta that are fit to the
         {\sc RAGE} output.  The black points denote the the SPH
         particles within the cosmological halo hosting the
         explosion. }
  \end{center}
\end{figure}

\section{Simulation Setup}
Here we describe the two simulations that we have carried out.  The
first is a 1-D radiation hydrodynamics calculation using the Los
Alamos National Laboratory RAGE code (Gittings et al.\ 2008) which
allows to track the propagation of the SN blast wave out to several
parsecs from the explosion site, deep within the host atomic cooling
halo.  For the second, we map the results of the first into a 3-D
cosmological simulation using the GADGET hydrodynamics code (Springel
et al.\ 2001; Springel \& Hernquist 2002).  Our use of these two
simulation codes for the phases of the SN in which they are most
well-suited to accurately model the explosion, from small (AU) scales
to large (kpc) scales, constitutes a significant improvement over
previous cosmological simulations of SN feedback.

\begin{figure*}
  \begin{center}
    \leavevmode
      \epsfxsize=7.in\epsfbox{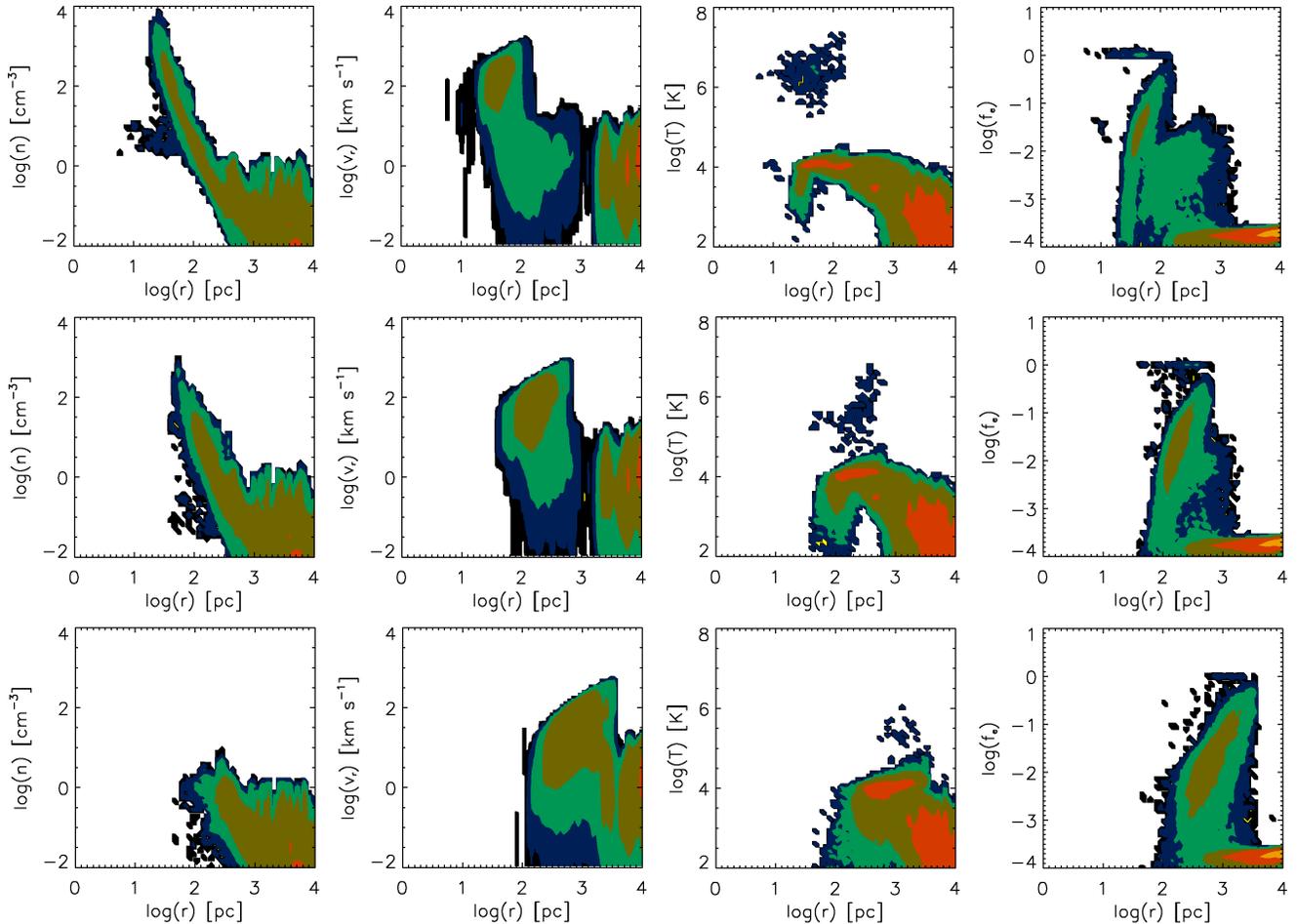}
       \caption{Properties of the gas in the vicinity of the SN at
         $100\,$kyr ({\it top panels}), $1\,$Myr ({\it middle panels})
         and $10\,$Myr ({\it bottom panels}) after the explosion of
         the central SMS.  From left to right, as functions of the
         distance from the explosion site, are the number density of
         hydrogen nuclei, outward radial velocity, gas temperature and
         free electron fraction.  Contours denote the distribution of
         the gas, with the mass fraction varying by an order of
         magnitude across contour lines.  The $\sim
         1000\,$km$\,$s$^{-1}$ shock completely disperses the dense
         ($n> 10\, $cm$^{-3}$) gas in the center of the host atomic
         cooling halo, but is only slightly decelerated even after
         propagating well beyond the $\sim 1\,$kpc virial radius of
         the halo.}
  \end{center}
\end{figure*}

\begin{figure}
  \begin{center}
    \leavevmode
      \epsfxsize=8.5cm\epsfbox{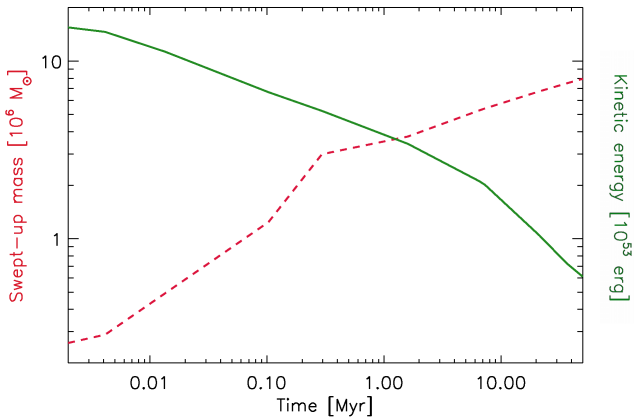}
       \caption{The mass of material swept up by the shock ({\it red
           dashed line}) and the total kinetic energy of the swept-up
         material ({\it green solid line}), as a function of time
         since the SN.  The majority of the kinetic energy is radiated
         away within the first $10^4\,$yr, and after $50\,$Myr all but
         $\sim1\,\%$ has been lost.  The majority of the material
         that is overtaken by the shock is swept-up after $1\,$Myr, with
         the total swept-up mass approaching $10^7\,$M$_{\odot}$.}
  \end{center}
\end{figure}

\begin{figure*}
  \begin{center} 
    \leavevmode
      \epsfxsize=7in\epsfbox{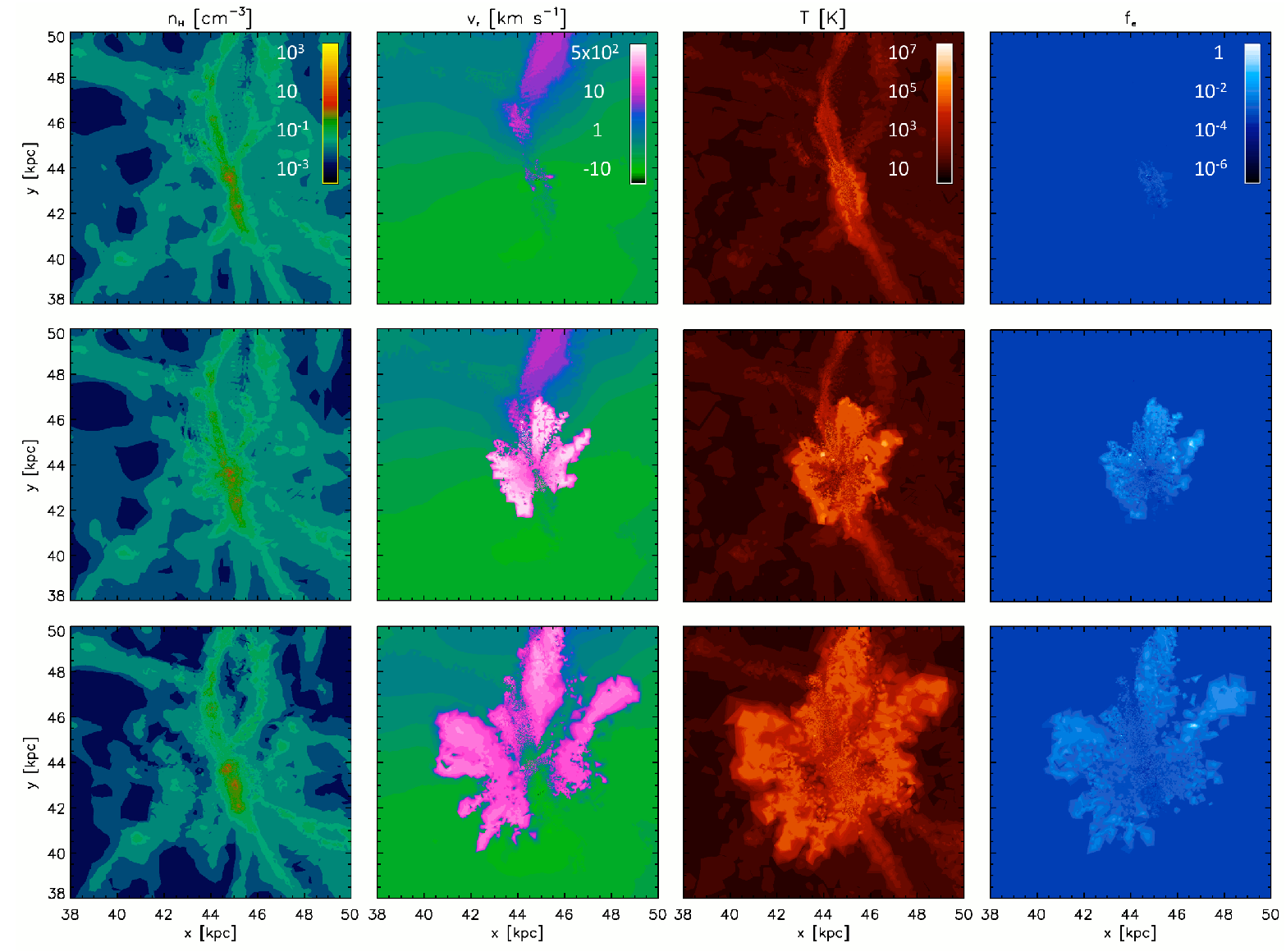}
       \caption{Properties of the gas in the vicinity of the SN at
         $1\,$Myr ({\it top panels}), $10\,$Myr ({\it middle panels})
         and $50\,$Myr ({\it bottom panels}) after the explosion.
         From left to right, are projections of the number density of
         hydrogen nuclei, outward radial velocity, gas temperature and
         free electron fraction, within a $400\,$pc comoving slice
         around the explosion site.  The lighter colors denote higher
         values for the quantities shown (see also Fig.~2).  The
         explosion is asymmetric, with the shock propagating more
         rapidly into the low-density voids, instead of into the
         high-density filaments, of the cosmological density field
         (see also Fig.~7).  While much of the swept-up mass is still
         outgoing at large velocities even after $50\,$Myr ({\it pink
           and white}), at this time metal-enriched gas is beginning
         to recollapse into the host halo where it will likely form
         second-generation stars (see Fig.~6).}
  \end{center}
\end{figure*}

\subsection{Stellar and Early Supernova Evolution}

For the SMS progenitor of the SN we adopt the $55,000\,$M$_{\odot}$
stellar model described in Whalen et al.\ (2012a), which was evolved
until the onset of explosion, using the {\it Kepler} code (Weaver et
al.\ 1978; Woosley et al.\ 2002).  The explosion was then followed
using {\it Kepler} and was confirmed (Chen et al., 2013) using the
CASTRO code (Almgren et al.\ 2010).  The explosion completely disrupts
the star and yields an explosion energy of 7.7 $\times$ 10$^{54}$ erg
(Heger et al.\ 2013).  The RAGE code is then used to simulate the SN
from the breakout of the shock from the surface of the star until the
blast wave has propagated through a circumstellar medium with a
density $\ge10^2\,$cm$^{-3}$ (and a density profile $\propto$
$r^{-2}$) out to several parsecs from the explosion site.  Up to this
point, the stellar evolution calculation and the simulation of the
early phases of the SN are the same as described in Whalen et
al.\ (2012a), to which we refer the reader for more detailed
discussion of the calclulations (see also Frey et al.\ 2013).

The blue curves in Figure 1 show the velocity (left panel) and density
(right panel) profiles of the SN ejecta at the end of the RAGE
simulation, at which point the shock has propagated out to $6\,$pc
from the explosion site.  The $55,000\,$M$_{\odot}$ of ejecta,
$23,000\,$M$_{\odot}$ of which is heavy elements produced during the
evolution and explosion of the progenitor, are traveling at almost
$10,000\,$km$\,$s$^{-1}$ outward from the center of the host atomic
cooling halo.  It is from this point that we map these velocity and
density profiles into a self-consistently evolved atomic cooling halo
in a much larger cosmological volume.

\subsection{Cosmological Blast Wave}

To simulate the subsequent evolution of the SN blast wave in the
appropriate cosmological environment, we map the velocity and density
profiles obtained from the smaller-scale RAGE simulation into the
center of a $4 \times 10^7\,$M$_{\odot}$ atomic cooling DM halo, the
type of which is expected to host the formation of SMSs in the early
universe.  The halo is identified in a $1\,$Mpc$^3$ (comoving)
cosmological volume which has been evolved from $z = 100$ down to $z
\simeq 15$ under the influence of a uniform, elevated H$_{\rm
  2}$-dissociating (Lyman-Werner; LW) radiation field, which prevents
the gas from cooling and is assumed to lead to formation of a single
SMS.  Further details of the cosmological simulation up to this point
are described in Johnson et al.\ (2011), who considered the impact of
the alternative end state of such a SMS, a rapidly accreting BH.

To map the output of the (Eulerian) RAGE simulation into the
(Lagrangian) smoothed particle hydrodynamics (SPH) GADGET simulation,
we assigned the central $460$ SPH particles, constituting the
$55,000\,$M$_{\odot}$ of gas within $\simeq 6\,$pc of the densest
particle in the halo, outward (radial) velocities and hydrogen number
densities so as to match those from the RAGE simulation.  The fits
that we obtain are shown in Fig.~1, with the inner $55,000\,$M$_{\odot}$
in SPH particles constituting the ejecta denoted by orange triangles
and the unpertured particles residing in the outskirts of the halo
denoted by black circles.  Whereas we fit the velocity profile very
well, due to the mapping from an Eulerian to Lagrangian code the
density profile is somewhat noisier.\footnote{Note in Fig.~1 that the
  density spike at the shock front in the RAGE output contains less
  mass than is contained in a single GADGET SPH particle, and so there
  are no particles representing this particular parcel of gas in the
  cosmological simulation.  This illustrates the fundamental
  difficulty in matching output from an Eulerian, adaptive mesh
  refinement code to an SPH code.}  Nonetheless, the basic features of
the density profile containing the vast majority of the mass are
represented, and the overall energy and momentum are also
well-matched.  From the initial conditions shown in Fig.~1 we restart
the cosmological SPH simulation, the results of which we present in
the next section.

Beyond mapping into it the blast profile of the SMS SN, we have chosen
to leave the gas in the host halo otherwise unchanged.  We have made
this simplifying choice, in light of the large uncertainties in the
radiative output of rapidly accreting SMSs, due to which it is unclear
how the radiation emitted during the brief ($\sim 2\,$Myr) lifetime of
the star will impact the medium within the host halo.  Even if the
star emits copious ionizing radiation, as main sequence Pop~III stars
are expected to do, it may be that the H~{\sc ii} region created by
the star is confined to the innermost regions of the halo (Johnson et
al.\ 2012; see also Hosokawa et al.\ 2012 on the possibility of even
less energetic radiation being emitted during the protostellar phase).
It is possible, on the other hand, that ionizing radiation is able to
escape out into the halo, if the accretion flow is highly anisotropic
(e.g., due to the presence of an accretion disk; e.g., McKee \& Tan
2008), or if accretion is intermittent (e.g., Clark et al.\ 2011;
Smith et al.\ 2011; Vorobyov et al.\ 2013), in which case radiation
could break out during periods of reduced accretion.

Whereas the small-scale structure of the interstellar gas in the
simulation we present here is subject to the limited resolution of the
cosmological simulation into which we place the expanding blast wave,
we note that at sub-resolution scales ($\la1\,$pc) it is possible that
a substantial amount of energy in the explosion is radiated away
(e.g., Kitayama \& Yoshida 2005; Whalen et al.\ 2008; de Souza et
al. 2011; Vasiliev et al. 2012).  We shall
address how such additional radiative losses would affect the dynamics
of the blast wave and the metal enrichment of the host halo and IGM in
future work.

\section{Results}
Here we present the results of our cosmological hydrodynamics
simulation, with particular attention paid to the dynamics of the
expanding blast wave and to the enrichment of the IGM by the 
metal-rich SN ejecta.

\subsection{Dynamics and Energetics}
The injection of almost $10^{55}$ erg at a single explosion site has a
dramatic impact on the host halo.  Figure~2 shows the properties of
the gas in the vicinity of the halo, as a function of the distance
from the explosion site at its center.  As the left-most panels show,
the blast results in the complete evacuation of high-density ($n$
$\ga10\,$cm$^{-3}$) gas within $10\,$Myr, with the material overtaken
by the shock being carried out beyond the virial radius of the halo
(at $r\simeq10^3\,$pc) at up to $\simeq 10^3\,$km$\,$s$^{-1}$.  This
material is also shock heated to temperatures up to $\sim 10^8\,$K,
resulting in its almost complete ionization, as shown in the
right-panels.  The gas, however, rapidly cools due to inverse Compton
scattering of CMB photons, H and He atomic line emission, and
bremsstrahlung, as also found (using the same GADGET code) in the
less-energetic ($10^{52}\,$erg) pair instability supernova (PSN)
explosion in a $2.5\times10^5\,$M$_\odot$ DM halo simulated by Greif et
al.\ (2007), as well as in the 1-D calculation of a very energetic
($10^{54}\,$erg) Pop III SN in a slightly less massive ($10^7
$M$_{\odot}$) DM halo presented in Kitayama \& Yoshida (2005).

As shown in Figure~3, most ($\simeq 90\,\%$) of the $7.7 \times
10^{54}\,$erg initially in the blast is radiated away via these
processes within $10^4\,$yr.  Nevertheless, the momentum of the blast
is conserved and the shock continues to propagate into the IGM,
sweeping up the majority of the mass after $\simeq 1\, $Myr.  By $50\,
$Myr, at which time $\simeq 99\,\%$ of the energy has been
radiated away, the shock has propagated out to $\simeq 5$--$10\,$kpc and
has swept up almost $10^7\,$M$_{\odot}$.

Figure~4 shows the properties of the gas in the vicinity of the
explosion site within a $400\,$pc (comoving) slice of the cosmological
volume, at $1\, $Myr, $10\, $Myr and $50\, $Myr after the explosion of
the SMS.  Comparing the radial velocity field (second column from the
left) to the cosmological density field (far left column), it is clear
that the blast wave propagates most rapidly into the low-density voids
while its progress is halted in the direction of the high-density
filaments, at the intersection of which lies the host halo.  Figs.~2
and 4 also show the same general trend that, at late times, the most
strongly shock-heated and highest-velocity material is located behind
the shock front several kpc from the explosion site.  The cooler
material within the ($\simeq 1\, $kpc) virial radius of the host halo is
able to begin recollapsing after $50\, $Myr.  As we discuss next, this gas
is likely to form second-generation stars that are enriched to fairly
high metallicities.

\begin{figure}
  \begin{center}
    \leavevmode
      \epsfxsize=8.7cm\epsfbox{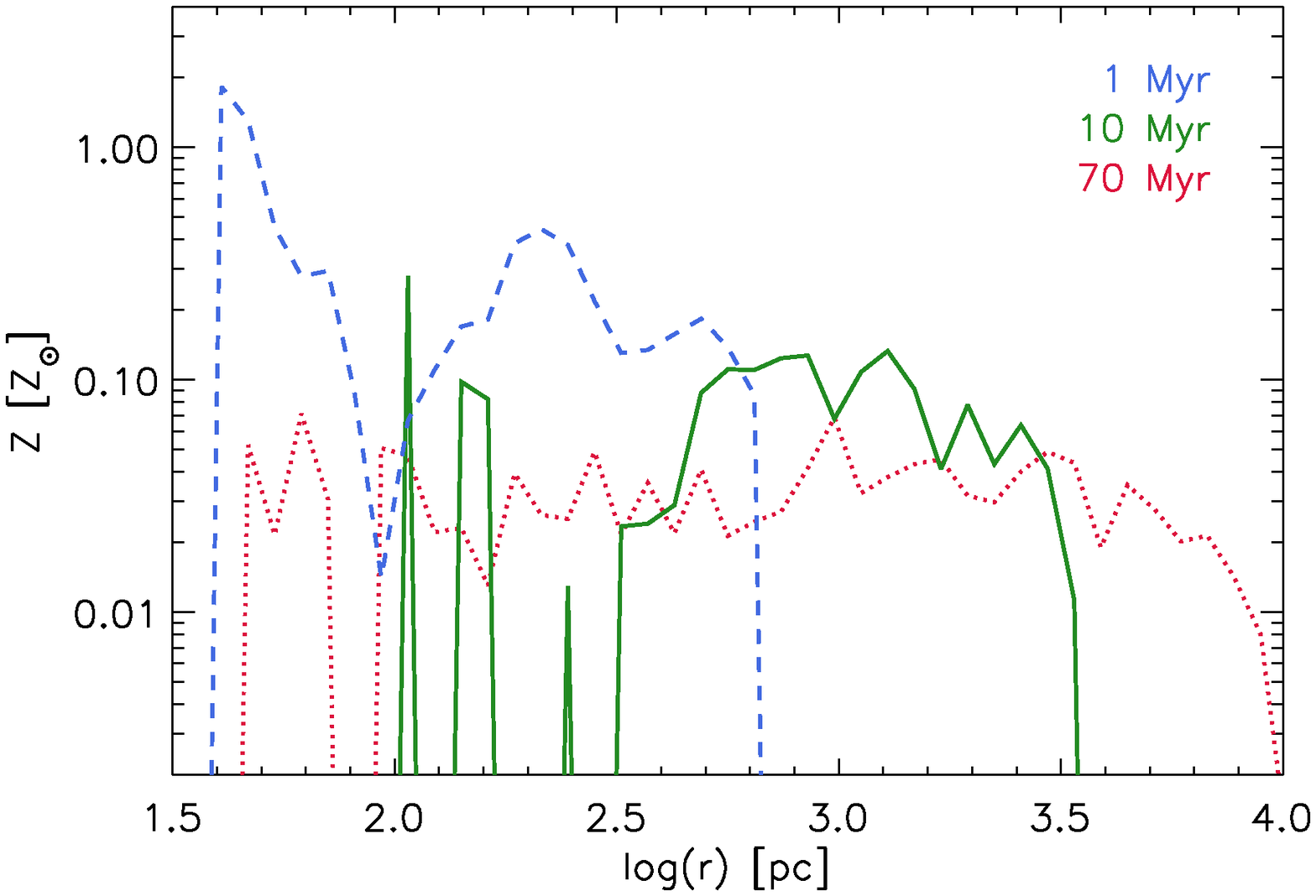}
       \caption{The spherically-averaged metallicity, $Z$, of the gas
         (in units of the solar metallicity Z$_{\odot}$), as a
         function of distance $r$ from the explosion site, at $1\,
         $Myr ({\it blue dashed line}), $10\, $Myr ({\it green solid
           line}) and $70\, $Myr ({\it red dotted line}).  As the
         metal-enriched gas is blown out of the halo, the metallicity
         in the interior region drops rapidly, while the IGM (outside
         the $\simeq 10^3\,$pc virial radius of the host halo) is
         enriched with metals out to continually larger radii.  After
         $70\, $Myr, the gas recollapsing into the host halo (see also
         Fig.~6) has a metallicity of $\simeq 0.05 \,$Z$_{\odot}$.}
  \end{center}
\end{figure}

\begin{figure}
  \begin{center}
    \leavevmode
      \epsfxsize=8.6cm\epsfbox{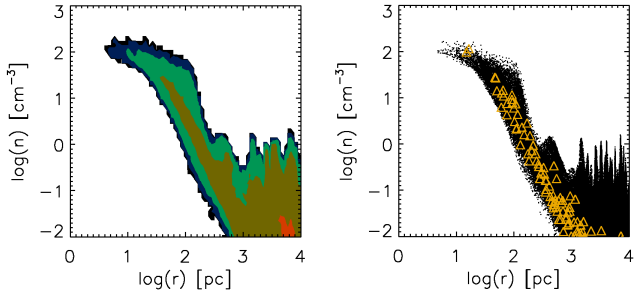}
       \caption{The hydrogen nuclei number density $n$, as a function
         of the distance $r$ from the explosion site, at $70\, $Myr
         after the SN.  As in Fig.~2, the contours in the left panel
         denote the distribution of the gas, the mass fraction varying
         by an order of magnitude across contour lines.  Individual
         SPH particles are shown in the right panel, with those
         constituting the metal-rich SN ejecta denoted by orange
         triangles, as in Fig.~1.  The highest-density gas is
         recollapsing into the halo which hosted the SN, where it will
         likely form second-generation stars.  While the majority of
         the metal-rich ejecta particles are in the low density
         intergalactic medium, the densest gas comprises some of them,
         suggesting that the second-generation stars will be highly
         metal-enriched. Many such stars would likely be long-lived
         and could still exhibit the chemical signature of SMS SNe
         today.}
  \end{center}
\end{figure}

\begin{figure*}
  \begin{center}
    \leavevmode
      \epsfxsize=7.1in\epsfbox{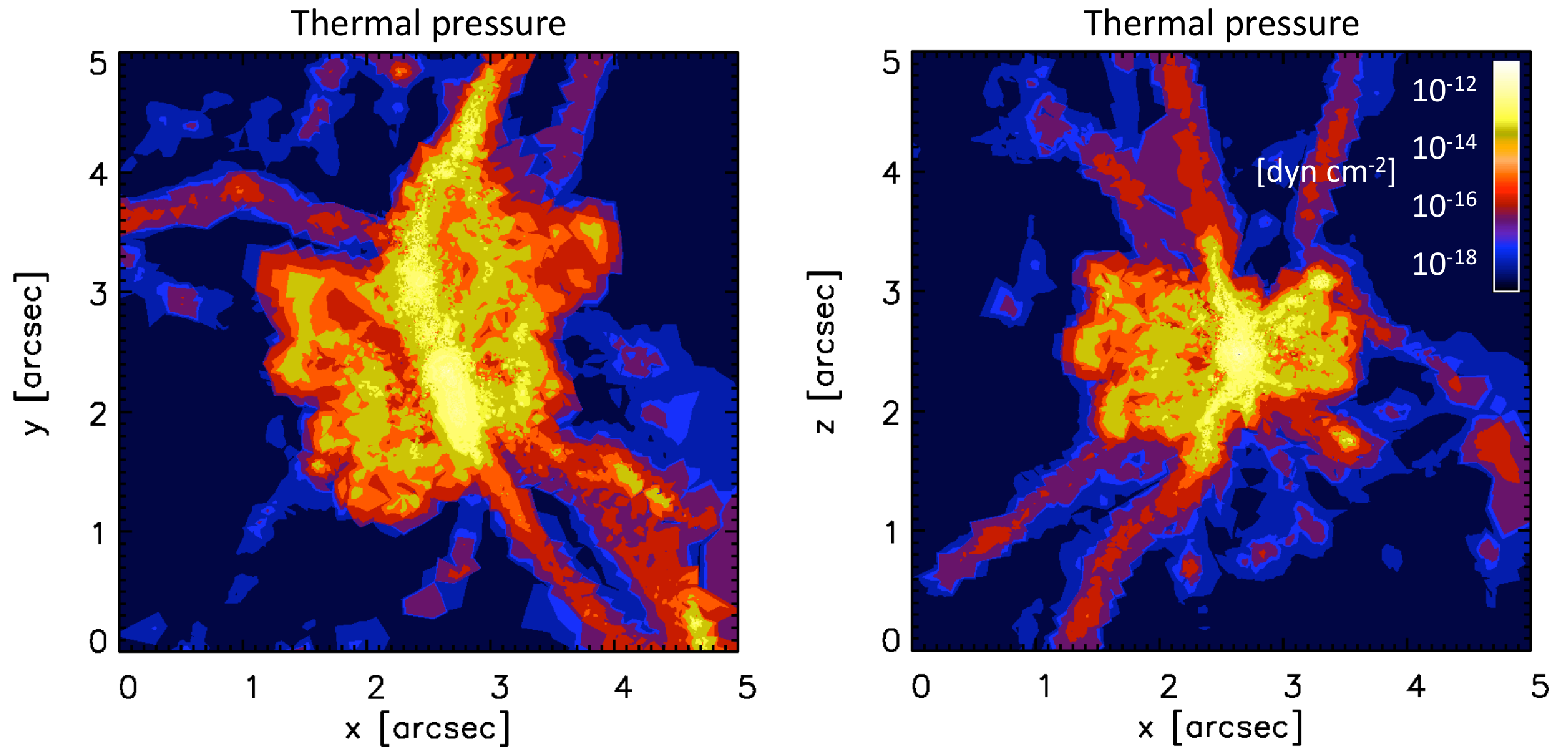}
       \caption{The SMS SN remnant 70 Myr after the explosion (at $z$
         $\simeq$ 13), as seen on the sky from two orthogonal viewing
         angles, with colors corresponding to the thermal pressure of
         the gas as shown.  Due to the inhomogeneous cosmological
         density field into which the blast wave propagates, the
         remnant takes on an irregular shape which can subtend up to
         $\sim 3$ arcsec, making it many times larger than most
         galaxies at this early epoch.  Indeed, it is likely that such
         remnants could envelop entire early star-forming galaxies
         (see Section 4).  SMS SNe may be detected in all-sky surveys,
         such as those planned for WFIRST and WISH.  At the center of
         this remnant, a cluster of metal-enriched second-generation
         stars is expected to form (see Fig.~6).}
  \end{center}
\end{figure*}

\subsection{Metal Enrichment and Second-Generation Star Formation}

As is also expected for less energetic PSNe from Pop~III stars with
masses of $\sim 200 $M$_{\odot}$ (e.g., Heger et al.\ 2003), a large
fraction of the ejecta from our SMS SN consists of heavy elements.  In
particular, $\simeq 23,000 \,$M$_{\odot}$ of newly-synthesized metals
are ejected in the explosion.  These heavy elements are mixed with the
primordial gas, enriching it to relatively high metallicity.

Figure~5 shows the average metallicity to which the gas in the
vicinity of the explosion site is enriched by the ejecta, after $1\,
$Myr, $10\, $Myr and $70\, $Myr.  As the blast wave propagates outward
into the low-density IGM gas at greater and greater radii becomes
enriched.  As metals are carried out of the host halo, metallicities
at the smallest radii fall.  After $70\, $Myr, however, the average
metallicity of the gas out to almost $10\, $kpc (physical) is enriched
to of the order of $10^{-2}\,$Z$_{\odot}$, and the densest gas which
is recollapsing into the host halo is enriched to $\simeq 0.05\,
$Z$_{\odot}$.  As shown in Figure~6, by $70\, $Myr the gas at the
center of the SN remnant, shown in Figure~7, has recollapsed to
densities of $n\sim10^2\,$cm$^{-3}$, significantly more dense than
the highest-density ($n$ $\simeq$10$\,$cm$^{-3}$) gas that remained
within the virial radius of the host halo $\sim 10\, $Myr after the SN.
Not only is the gas recollapsing after $70\, $Myr, but, as shown in the
right panel of Fig.~6, a portion of the SN ejecta is entrained in this
gas.  If this ejecta is well-mixed with the dense primordial gas, then
we expect its metallicity to be $\simeq 0.05\, $Z$_{\odot}$, as
indicated in Fig.~5.

Gas enriched to such a high metallicity is predicted to readily
fragment into low-mass stars (e.g., Bromm et al.\ 2001; Santoro \&
Shull 2006; Schneider et al.\ 2006), even in the presence of the
elevated LW background radiation field expected in regions of the
universe where SMSs form (e.g., Omukai et al.\ 2008).  Therefore, we
expect that second-generation stars would form from this SMS
SN-enriched gas, and that a large fraction of these would have masses
low enough ($\la 0.8 \,$M$_{\odot}$) that they could still be present
in the Galaxy today.  While the nucleosynthetic signature of very
massive Pop III pair-instability SNe (i.e., $140$--$260\,
$M$_{\odot}$) has yet to be uncovered in extremely metal-poor stars
(Cayrel et al.\ 2004; Beers \& Christlieb 2005; Frebel et al 2005; Lai
et al 2008; Joggerst et al 2010; Joggerst \& Whalen 2011), it may have
been found in high-redshift damped Lyman alpha absorbers (Cooke et
al.\ 2011), and a number of metal-poor stars in a recent extension to
the SEGUE survey have now been selected for spectroscopic followup on
suspicion that they too may harbor this pattern (Ren et al.\ 2012).
Furthermore, most stars forming in the ashes of very massive
primordial SNe were likely enriched to metallicities above those
targeted by surveys of metal-poor stars to date (Karlsson et
al.\ 2008), and our simulations predict that second-generation stars
formed from gas enriched by SMS SNe will have metallicities above
$10^{-2}\,$Z$_{\odot}$, which are also above this threshold.  We
conclude that, while SMS SNe are almost certainly rare events, their
chemical signature might well be found in some of the ancient but not
very metal-poor stars inhabiting our Galaxy today.

\section{Discussion and Conclusions}

We have carried out a multi-scale simulation of the explosion of a
$55,000\, $M$_{\odot}$ SMS, which, with an energy approaching
$10^{55}\,$erg, is among the biggest explosions in the history of the
universe.  With our multi-code approach, we have captured
self-consistently, and for the first time, the essential radiation
hydrodynamic features of the explosion at early times and the
interaction of the blast wave with its host protogalaxy and with the
IGM at later times.

Although the atomic cooling halos expected to host the formation of
SMSs are at least two orders of magnitude more massive than the DM
halos in which the first primordial stars are expected to form, we
have found that these SMS SNe are energetic enough to completely
evacuate them of dense gas.  The metal-enriched ejecta is dispersed
well beyond the $\sim 1\, $kpc virial radius of the host halo, out to
$\sim 5$--$10\, $kpc into the low-density IGM, after $\sim 50\, $Myr.
By this time, $\sim 99\,\%$ of the kinetic energy in the explosion has
been radiated away; nonetheless, the expansion of the shock into the
IGM continues even at these late times, as the remaining kinetic
energy is still comparable to that expected for a $\sim 200
\,$M$_{\odot}$ Pop~III PSN.

Because of the deep potential well of the halo and ongoing accretion
from filaments, after $70\, $Myr a fraction of the metal-enriched gas
in the SN remnant, shown in Fig.~7, has recollapsed to high densities.
Given the relatively high metallicity of this gas ($\sim 0.05
\,$Z$_{\odot}$),\footnote{Previous cosmological simulations of metal
  enrichment by massive Pop~III stars have shown second-generation
  star-forming gas to have typical metallicities of $\ga 10^{-3}
  \,$Z$_{\odot}$ (e.g., Wise \& Abel 2008; Greif et al.\ 2010; Ritter
  et al.\ 2012; Vasiliev et al.\ 2012; see also Wise et al.\ 2012;
  Safranek-Shrader et al. 2013).
  Incidentally, we note that Greif et al.\ (2007) arrived at
  comparable results by tracking metal enrichment using a much smaller
  number of SPH particles than we have used to track it here.  Thus,
  we expect that our results with regard to metal enrichment should be
  broadly consistent with what would be found using other approaches.}
it is most likely that it will fragment vigorously and form a cluster
of second-generation stars.  Enriched to this metallicity, any of
these stars which survive to the present-day would exhibit
metallicities much higher than the most metal-poor stars that have
been found in surveys of the Galactic halo.  Indeed, although SMS SN
are likely rare events, it is possible that their chemical signature
could be found in very old, relatively high-metallicity Pop~II stars
which inhabit the Milky Way today (or in present-day dwarf galaxies;
Frebel \& Bromm 2012).  As we expect SMS SNe to produce
very little $^{56}$Ni (Heger et al.\ 2013), their chemical signature
may be distinguishable from those of most very massive (i.e.,
$140$--$260 \,$M$_{\odot}$) Pop~III explosions (PSNe), which can
produce iron-group elements (Heger \& Woosley 2002).  Similar to these
PSNe, however, SMS SNe would not make any \textsl{s}-process or
\textsl{r}-process contributions.

Finding the chemical signature of SMS SNe in low-mass, long-lived
stars (or in the IGM; see e.g., Cooke et al.\ 2011) would be one way
of verifying that such exotic events occurred in the early universe.
Another possibility is that these gargantuan explosions could be found
in all-sky surveys such as those planned for the {\it Wide-field
  Infrared Survey Telescope} (WFIRST) and the {\it Wide-field Imaging
  Surveyor for High-redshift} (WISH), as shown recently by Whalen et
al.\ (2012a).\footnote{We note that neutrino emission from these
  explosions (produced as discussed in e.g., Kistler \& Beacom 2006;
  Yamazaki 2009; Morlino et al.\ 2009; Yuan et al.\ 2010) might also
  be detectable.}  Given that SMSs are expected to form in regions
subjected to a large flux of LW radiation from nearby (within $\sim
10\, $kpc; Dijkstra et al.\ 2008; Agarwal et al.\ 2012) star-forming
galaxies, detection of SMS SNe would pinpoint locations on the sky
where rapidly-forming first galaxies could be found in follow-up
observations by the {\it James Webb Space Telescope}.  Indeed, these
explosions are so large that the SN remnants they leave behind, as
shown in Fig.~7, are likely to be many times larger than, and in fact
are likely to envelop, any such neighboring star-forming galaxies.

Could later stages of SMS SNe be detected by other means?  They
might appear in the radio at $21\, $cm.  Meiksin \& Whalen (2013) 
recently found that synchrotron emission from hypernovae in 
relatively dense media will be visible at $21\, $cm to existing radio 
facilities such as eVLA and eMERLIN in  addition to the 
{\it Square Kilometer Array} (SKA).  With their much higher energies 
and similar circumstellar densities, SMS SNe may be much 
brighter in the radio and detectable in all-sky surveys in spite of 
their small numbers.  We are now calculating the radio signatures
of SMS SNe in $z\sim 15$ protogalaxies.

Likewise, Oh et al.\ (2003) and Whalen et al.\ (2008) examined 
the potential imprint  of Pop III SNe on the cosmic microwave 
background (CMB) via the Sunyayev-Zeldovich (SZ) effect.  
They found that a population of $140$--$260\, $M$_{\odot}$ PSNe might 
impose excess power on the CMB on small scales, but with two 
caveats.  First, the explosions must occur in large H~{\sc ii} regions 
because the SN shock must still be hot by the time it encloses a 
large volume of CMB photons.  Explosions in dense media 
dissipate too much heat to have an appreciable SZ signature or 
upscatter many CMB photons.  Second, although a population
of Pop III SNe might collectively impose features on the CMB
individual remnants achieve radii just below the current resolution
of {\it Atacama Cosmology Telescope} or the {\it South Pole Telescope}.
Our models explode in dense environments and at redshifts at
which inverse Compton cooling losses are lower than for the
first SNe, but they eventually reach radii that would allow them to 
be resolved by current instruments.  We are currently evaluating the 
SZ signatures of SMS SNe.

\section*{Acknowledgements}

This work was supported by the U.S. Department of Energy through the
LANL/LDRD Program, and JLJ acknowledges the support of a LDRD
Director's Postdoctoral Fellowship at Los Alamos National Laboratory.
The RAGE and GADGET simulations were carried out on the LANL
Institutional Computing clusters Pinto and Mustang, respectively.  DJW
acknowledges support from the Baden-W\"{u}rttemberg-Stiftung by
contract research via the programme Internationale Spitzenforschung II
(grant P- LS-SPII/18).  AH and KC were supported by the US DOE Program
for Scientific Discovery through Advanced Computing (SciDAC;
DE-FC02-09ER41618), by the US Department of Energy under grant
DE-FG02-87ER40328, by the Joint Institute for Nuclear Astrophysics
(JINA; NSF grant PHY08-22648 and PHY110-2511).  AH acknowledges
support by an ARC Future Fellowship (FT120100363) and a Monash
University Larkins Fellowship.  KC was supported by a KITP/UCSB
Graduate Fellowship and by a UMN Stanwood Johnston Fellowship.  The
authors thank Avery Meiksin for helpful discussion.  Work at LANL was
done under the auspices of the National Nuclear Security
Administration of the U.S. Department of Energy at Los Alamos National
Laboratory under Contract No. DE-AC52-06NA25396.

\newpage

\bibliographystyle{apj}

\begin{thebibliography}{199}

\bibitem[2(2000)]{}Abel, T., Bryan, G.~L., Norman, M.~L. 2002, Sci,
  295, 93
\bibitem[2(2000)]{}Almgren, A.~S., et al. 2010, ApJ, 715, 1221
\bibitem[2(2000)]{}Agarwal, B., Khochfar, S., Johnson, J.~L.,
  Neistein, E., Dalla Vecchia, C., Livio, M. 2012, MNRAS, 425, 2854
\bibitem[2(2000)]{}Agarwal, B., Davis, A.~J., Khochfar, S., Natarajan,
  P., Dunlop, J.~S. 2013, MNRAS, submitted (arXiv:1302.6996)
\bibitem[2(2000)]{a1}Alvarez, M.~A., Wise, J.~H., Abel, T. 2009, ApJ, 701, L133
\bibitem[2(2000)]{a1}Appenzeller, I., Fricke, K. 1972, A\&A, 21, 285
\bibitem[2(2000)]{}Begelman M.~C. 2010, MNRAS, 402, 673
\bibitem[2(2000)]{}Begelman M.~C., Volonteri M., Rees M.~J. 2006, MNRAS, 370, 289
\bibitem[2(2000)]{}Beers, T.~C., Christlieb, N. 2005, ARA\&A, 43, 531
\bibitem[2(2000)]{}Bond, J.~R., Arnett, W.~D., Carr, B.~J. 1984, ApJ, 280, 825
\bibitem[2(2000)]{}Bromm, V., Ferrara, A., Coppi, P.~S., Larson, R.~B. 2001, MNRAS, 328, 969
\bibitem[2(2000)]{}Bromm, V., Loeb, A. 2003, ApJ, 596, 34
\bibitem[2(2000)]{}Bromm V., Larson R.~B. 2004, ARA\&A, 42, 79
\bibitem[2(2000)]{}Cayrel, R., et al. 2004, A\&A, 416, 1117
\bibitem[2(2000)]{}Choi, J.-H., Shlosman, I., Begelman, M.~C. 2013,
  ApJ, submitted (arXiv:1304.1369)
\bibitem[2(2000)]{}Clark, P.~C., Glover, S.~C.~O., Smith, R.~J., Greift, T.~H., Klessen, R.~S., Bromm, V. 2011, Sci, 331, 1040
\bibitem[2(2000)]{}Cooke, R., Pettini, M., Steidel, C.~C., Rudie, G.~C., Jogenson, R.~A. 2011, MNRAS, 412, 1047
\bibitem[2(2000)]{}de Souza, R.~S., Rodrigues, L.~F.~S., Ishida,
  E.~E.~O., Opher, R. 2011, MNRAS, 415, 2969
\bibitem[2(2000)]{}de Souza, R.~S., Ishida, E.~E.~O., Johnson, J.~L.,
  Whalen, D.~J., Mesinger, A. 2013, MNRAS, submitted (arXiv:1306.4984)
\bibitem[2(2000)]{}Dijkstra, M., Haiman, Z., Mesinger, A., Wyithe, J.~S.~B. 2008, MNRAS, 391, 1961
\bibitem[2(2000)]{}Fan, X., et al. 2006, AJ, 131, 1203 
\bibitem[2(2000)]{}Fowler, W.~A., Hoyle, F. 1964, ApJS, 9, 201
\bibitem[2(2000)]{}Frebel, A., et al. 2005, Nat, 434, 871
\bibitem[2(2000)]{}Frebel, A., Bromm, V. 2012, ApJ, 759, 115
\bibitem[2(2000)]{}Frey, L., Even, W., Whalen, D.~J., et al. 2013,
  ApJS, 204, 16
\bibitem[2(2000)]{}Fryer, C.~L., Heger, A. 2011, AN, 332, 408
\bibitem[2(2000)]{}Fuller, G.~M., Woosley, S.~E., Weaver, T.~A. 1986, ApJ, 307, 675
\bibitem[2(2000)]{}Fuller, G.~M., Shi, X. 1998, ApJ, 502, L5
\bibitem[2(2000)]{}Gittings, M., et al. 2008, CS\&D, 1, 5005
\bibitem[2(2000)]{}Greif T.~H., Johnson J.~L., Bromm V., Klessen R.~S. 2007, ApJ, 670, 1
\bibitem[2(2000)]{}Greif T.~H., Glover, S.~C.~O., Bromm, V., Klessen,
  R.~S. 2010, ApJ, 716, 510
\bibitem[2(2000)]{}Greif T.~H., Springel, V., White, S.~D.~M., Glover,
  S.~C.~O., Clark, P.~C., Smith, R.~J., Klessen, R.~S., Bromm,
  V. 2011, ApJ, 737, 75
\bibitem[2(2000)]{b}Heger, A., Woosley, S.~E. 2002, ApJ, 567, 532
\bibitem[2(2000)]{b}Heger, A., et al. 2013, in prep
\bibitem[2(2000)]{b}Hosokawa, T., Omukai, K., Yorke, H.~W. 2012, ApJ, submitted (arXiv:1203.2613)
\bibitem[2(2000)]{b}Hummel, J.~A., Pawlik, A.~H., Milosavljevi{\' c},
  M., Bromm, V. 2012, ApJ, 755, 72
\bibitem[2(2000)]{b}Iben, I. 1963, ApJ, 138, 1090
\bibitem[2(2000)]{b}Inayoshi, K., Omukai, K. 2012, MNRAS, 422, 2539
\bibitem[2(2000)]{b}Inayoshi, K., Hosokawa, T., Omukai, K. 2013, MNRAS, accepted (arXiv:1302.6065)
\bibitem[2(2000)]{b}Jeon, M., Pawlik, A.~H., Greif, T.~H., Glover, S.~C.~O., Bromm, V., Milosavljevi{\' c}, M., Klessen, R.~S. 2012, ApJ, 754, 34
\bibitem[2(2000)]{b}Joggerst, C.~C., Whalen, D.~J. 2011, ApJ, 728, 129
\bibitem[2(2000)]{b}Joggerst, C.~C., Almgren, A., Bell, J., Heger, A.,
  Whalen, D.~J., Woosley, S.~E. 2010, ApJ, 709, 11

\bibitem[2(2000)]{b}Johnson, J.~L., Dalla Vecchia, C., Khochfar, S. 2013, MNRAS, 428, 1857
\bibitem[2(2000)]{b}Johnson, J.~L., Khochfar, S., Greif, T.~H., Durier, F. 2011, MNRAS, 410, 919
\bibitem[2(2000)]{}Johnson, J.~L., Whalen, D.~J., Fryer, C.~L., Li, H. 2012b, ApJ, 750, 66
\bibitem[2(2000)]{}Johnson, J.~L., Whalen, D.~J., Li, H., Holz, D.~E. 2012a, ApJ, submitted (arXiv:1211.0548)
\bibitem[2(2000)]{}Karlsson, T., Johnson, J.~L., Bromm, V. 2008, ApJ, 679, 6
\bibitem[2(2000)]{}Kistler, M.~D., Beacom, J.~F. 2006, PhRvD, 74, 063007
\bibitem[2(2000)]{}Kitayama T., Yoshida N. 2005, ApJ, 630, 675
\bibitem[2(2000)]{}Komatsu, E., et al. 2011, ApJS, 192, 18
\bibitem[2(2000)]{b}Lai, D.~K., Bolte, M., Johnson, J.~A., Lucatello,
  S., Heger, A., Woosley, S.~E. 2008, ApJ, 681, 1524
\bibitem[2(2000)]{b}Latif, M.~A., Schleicher, D.~R.~G., Schmidt, W.,
  Niemeyer, J. 2013a, MNRAS, 430, 588
\bibitem[2(2000)]{b}Latif, M.~A., Schleicher, D.~R.~G., Schmidt, W.,
  Niemeyer, J. 2013b, MNRAS, submitted (arXiv:1304.0962)
\bibitem[2(2000)]{b}Linke, F., Font, J.~A., Janka, H.-T., M{\"u}ller, E., Papadopoulos, P. 2001, A\&A, 376, 568
\bibitem[2(2000)]{b}Lodato, G., Natarajan, P. 2006, MNRAS, 371, 1813
\bibitem[2(2000)]{b}McKee, C.~F., Tan, J.~C. 2008, ApJ, 681, 771
\bibitem[2(2000)]{b}Meiksin, A., Whalen, D.~J. 2013, MNRAS, 430, 2854
\bibitem[2(2000)]{b}Milosavljevi{\' c}, M., Bromm, V., Couch, S.~M.,
  Oh, S.~P. 2009, ApJ, 698, 766
\bibitem[2(2000)]{}Montero, P.~J., Janka, H.-T., M{\" u}ller, E. 2012,ApJ, 749, 37
\bibitem[2(2000)]{}Morlino, G., Blasi, P., Amato, E. 2009,
  Astropart. Phys., 31, 376
\bibitem[2(2000)]{}Mortlock, D.~J., et al. 2011, Nat, 474, 616
\bibitem[2(2000)]{b}Natarajan, P., Volonteri, M. 2012, MNRAS, 422, 2051
\bibitem[2(2000)]{b}Oh, S.~P., Cooray, A., Kamionkowski, M. 2003,
  MNRAS, 342, L20
\bibitem[2(2000)]{}Omukai, K., Schneider, R., Haiman, Z. 2008, ApJ, 686, 801
\bibitem[2(2000)]{}O'Shea B.~W., Norman M.~L. 2008, ApJ, 673, 14
\bibitem[2(2000)]{a1}Pan, T., Kasen, D., Loeb, A. 2012, MNRAS, 422, 2701
\bibitem[2(2000)]{a1}Park, K., Ricotti, M. 2012, ApJ, 747, 9
\bibitem[2(2000)]{a1}Pelupessy, F.~I., Di Matteo, T., Ciardi, B. 2007, ApJ, 665, 107--119
\bibitem[2(2000)]{b}Petri, A., Ferrara, A., Salvaterra, R. 2012, MNRAS, accepted (arXiv:1202.3141)
\bibitem[2(2000)]{b}Planck Collaboration, A\&A submitted (arXiv:1303.5076)
\bibitem[2(2000)]{}Regan J.~A., Haehnelt M.~G. 2009, MNRAS, 396, 343
\bibitem[2(2000)]{}Ren, J., Christlieb, N., Zhao, G. 2012, RAA, 12, 1637
\bibitem[2(2000)]{}Ritter, J.~S., Safranek-Shrader, C., Gnat, O.,
  Milosavljevi{\' c}, M., Bromm, V. 2012, ApJ, 761, 56
\bibitem[2(2000)]{}Safranek-Shrader, C., Milosavljevi{\' c}, M.,
  Bromm, V. 2013, MNRAS, submitted (arXiv:1307.1982)
\bibitem[2(2000)]{}Santoro, F., Shull, J.~M. 2006, ApJ, 643, 26
\bibitem[2(2000)]{}Scannapieco, E., Madau, P., Woosley, S., Heger, A.,
  Ferrara, A. 2005, ApJ, 633, 1031
\bibitem[2(2000)]{}Schleicher, D.~R.~G., Palla, F., Ferrara, A.,
  Galli, D., Latif, M. 2013, A\&A, submitted (arXiv:1305.5923)
\bibitem[2(2000)]{}Schneider, R., Omukai, K., Inoue, A.~K., Ferrara,
  A. 2006, MNRAS, 369, 825
\bibitem[2(2000)]{b}Sethi, S., Haiman, Z., Pandey, K. 2010, ApJ, 721, 615
\bibitem[2(2000)]{b}Shang, C., Bryan, G.~L., Haiman, Z. 2010, MNRAS, 402, 1249
\bibitem[2(2000)]{b}Shapiro, S.~L. 2005, ApJ, 620, 59
\bibitem[2(2000)]{}Shapiro, S.~L., Teukolsky, S.~A. 1979, ApJ, 234, L177
\bibitem[2(2000)]{}Smith, R.~J., Glover, S.~C.~O., Clark, P.~C.,
  Greif, T.~H., Klessen, R.~S. 2011, MNRAS, 414, 3633
\bibitem[2(2000)]{}Spaans, M., Silk, J. 2006, ApJ, 652, 902
\bibitem[2(2000)]{}Springel V., Yoshida N., White S.~D.~M., 2001,
  NewA, 6, 79
\bibitem[2(2000)]{}Springel V., Hernquist, L. 2002, MNRAS, 333, 649
\bibitem[2(2000)]{}Tanaka, M., Moriya, T.~J., Yoshida, N., Nomoto,
  K. 2012, MNRAS, 422, 2675
\bibitem[2(2000)]{}Tanaka, M., Moriya, T.~J., Yoshida, N. 2013, MNRAS,
  submitted (arXiv:1306.3743)
\bibitem[2(2000)]{}Van Borm, C., Spaans, M. 2013, A\&A, submitted (arXiv:1304.4057)
\bibitem[2(2000)]{}Vasiliev, E.~O., Vorobyov, E.~I., Matvienko, E.~E.,
  Razoumov, A.~O., Shchekinov, Y.~A. 2012, ARep, 56, 895
\bibitem[2(2000)]{}Volonteri, M., Rees, M. 2006, ApJ, 650, 669
\bibitem[2(2000)]{}Volonteri, M. 2012, Sci, 337, 544
\bibitem[2(2000)]{}Volonteri, M., Begelman, M.~C. 2010, MNRAS, 409, 1022
\bibitem[2(2000)]{}Vorobyov, E.~I., DeSouza, A.~L., Basu, S. 2013,
  ApJ, submitted (arXiv:1303.3622)
\bibitem[2(2000)]{}Weaver, T.~A., Zimmerman, G.~B., Woosley,
  S.~E. 1978, ApJ, 225, 1021
\bibitem[2(2000)]{}Whalen D., van Veelen B., O'Shea B.~W., Norman M.~L. 2008, ApJ, 682, 49
\bibitem[2(2000)]{}Whalen, D.~J., Heger, A., Chen, K.-J., Even, W.,
  Fryer, C.~L., Stiavelli, M., Xu, H., Joggerst, C.~C. 2012a, ApJ, submitted (arXiv:1211.1815)
\bibitem[2(2000)]{}Whalen, D.~J., Abel, T., Norman, M.~L. 2004, ApJ, 610, 14
\bibitem[2(2000)]{}Whalen, D.~J., et al. 2013a, ApJ, 768, 195 
\bibitem[2(2000)]{}Whalen, D.~J., et al. 2012b, ApJ, submitted (arXiv:.1211.4979)
\bibitem[2(2000)]{}Whalen, D.~J., et al. 2013b, ApJ, 768, 95
\bibitem[2(2000)]{}Whalen, D.~J., et al. 2013c, ApJL, 762, 6 
\bibitem[2(2000)]{}Whalen, D.~J., et al. 2013d, ApJ, accepted (arXiv:1305.6966)
\bibitem[2(2000)]{}Willott, C.~J., McLure, R.~J., Jarvis, M.~J., 2003, ApJ, 587, L15
\bibitem[2(2000)]{}Wise, J.~H., Abel, T. 2007, ApJ, 671, 1559
\bibitem[2(2000)]{}Wise, J.~H., Turk, M.~J., Abel, T. 2008, ApJ, 682, 745
\bibitem[2(2000)]{}Wise, J.~H., Abel, T. 2008, ApJ, 685, 40
\bibitem[2(2000)]{}Wise, J.~H., Turk, M.~J., Norman, M.~L., Abel,
  T. 2012, ApJ, 745, 50
\bibitem[2(2000)]{}Wolcott-Green, J., Haiman, Z., Bryan, G.~L. 2011, MNRAS, 418, 838
\bibitem[2(2000)]{}Woosley, S.~E., Heger, A., Weaver, T.~A. 2002,
  RevMP, 74, 1015
\bibitem[2(2000)]{}Yamazaki, R., Kohri, K., Katagiri, H. 2009, A\&A,
  495, 9
\bibitem[2(2000)]{}Yoshida, N., Omukai, K., Hernquist, L. 2008, Sci,
  321, 669
\bibitem[2(2000)]{}Yuan, Q., Yin, P., Bi, X. 2010, arXiv:1010.1901

\end{thebibliography}

\end{document}